\documentclass[useAMS,usenatbib]{mn2e}

\usepackage{booktabs}
\usepackage{graphicx, subfigure}

\def\apj{ApJ}%
\def\apjl{ApJ}%
\def\apjs{ApJS}%
\def\aap{A\&A}%
\def\aaps{A\&AS}%
\def\mnras{MNRAS}%
\def\na{New A}%
\def\nar{New A Rev.}%
\def\pasp{PASP}%
\def\pasj{PASJ}%
\def\nat{Nature}%
\def\bain{Bull.~Astron.~Inst.~Netherlands}%
\title[Chemical abundances of the secondary star in the Neutron
star X-ray binary Cygnus X-2]{Chemical abundances of the
secondary star in the Neutron star X-ray binary Cygnus X-2}
\author[L. Su\'arez-Andr\'es, J.I. Gonz\'alez Hern\'andez, G.
Israelian, J. Casares and R. Rebolo ]{L.
Su\'arez-Andr\'es$^{1,2}$\thanks{E-mail: lsuarez@iac.es}, J.I.
Gonz\'alez Hern\'andez$^{1,2}$, G. Israelian$^{1,2}$, J.
Casares$^{1,2}$ \\ \newauthor
and R. Rebolo$^{1,2,3}$ \\
$^{1}$Instituto de Astrof\'isica de Canarias, E-38205 La Laguna,
Tenerife, Spain\\
$^{2}$Depto. Astrof\'isica, Universidad de La Laguna (ULL),
E-38206 La Laguna, Tenerife, Spain\\
$^{3}$Consejo Superior de Investigaciones Cient\'ificas, Spain
}

\begin{document}

\date{Accepted 2014 December 8. Received 2014 August 4}

\pagerange{\pageref{firstpage}--\pageref{lastpage}} 

\pubyear{2014}

\maketitle

\label{firstpage}

\begin{abstract}
We present UES@WHT high-resolution spectra of the low-mass X-ray binary (LMXB) Cygnus X-2. We have derived the stellar parameters of the secondary star using $\chi^{2}$ minimisation procedure, and taking into account any possible veiling from the accretion disk. We determine a metallicity higher than solar ([Fe/H]$=0.27\pm0.19$), as seen also in the neutron star X-ray binary Centaurus X-4.
The high-quality of the secondary's spectrum allow us to determine the chemical abundances of O, Mg, Si, Ca, S, Ti, Fe and Ni. We found that some $\alpha$-elements (Mg, Si, S, Ti) are enhanced, consistent with a scenario of contamination of the secondary star during the supernova event.
Surprisingly oxygen appears to be under-abundant, whereas enhanced abundances of Fe and Ni are measured. Assuming that these abundances come from matter that has been processed in the SN and then captured by the secondary star, we explore different SN explosion scenarios with diverse geometries.
A non-spherically symmetric SN explosion, with a low mass cut,  seems to reproduce better the observed abundance pattern of the secondary star compared to the spherical case.

\end{abstract}


\section{Introduction}

Cygnus X-2 (V1341 Cygni) is one of the few low-mass X-ray binaries (LMXB) in which the spectrum of the non-degenerate star is visible, contributing about 50$\% $ of the total flux. 
Cygnus X-2, with a mass function of $f(M) = 0.66 \pm 0.03$~\citep{cas98,cas10}, is also one of the brightest and most massive systems known to date, with a neutron star of $M_{\rm NS}=1.71\pm0.21M_{\odot}$. A mass ratio of $q=M_{2}/M_{\rm NS} = 0.34 \pm 0.02$ was determined from the measurement of the rotational velocity of the secondary star, $v \sin i= 34.6 \pm 0.1$~km~s$^{-1}$~\citep{cas10}.

The chemical abundances of secondary stars in black hole and neutron star X-ray binaries have been studied for several systems: Nova Scorpii 1994~\citep{isr99,gon08a}, A0620-00~\citep{gon04}, Centaurus X-4~\citep{gon05a,cas07}, XTE J1118+480~\citep{gon06,gon08b}, V404 Cygni~\citep{gon11}, and V4641 Sagittarii~\citep{oro01,sad06}. 
The metallicities of these binary systems are all similar or higher than solar regardless of their location with respect to the Galactic plane. In addition, the above authors have taken into account different scenarios of pollution from supernova (SN) and more energetic hypernova (HN) ejecta on the photospheric abundances of the secondary star. 	

Our reference for this work is Centaurus X-4, a neutron star X-ray binary like Cygnus X-2. \citet{gon05a} found in the secondary star of Centaurus X-4 a supersolar metallicity ([Fe/H]$\sim0.23\pm0.10$), as well as supersolar abundances of other important elements, like Ca, Ti or Al. These supersolar abundances may indicate that the mass-cut, i.e. the initial mass of the compact object right after the SN/HN explosion, is lower than in the case of black hole X-ray binaries (BHXBs).

In this paper, we use high-resolution spectra to derive the stellar parameters and chemical abundances of the secondary star in the neutron star X-ray binary (NSXB) Cygnus X-2 with the aim of obtaining information about its formation and evolution, the properties of the progenitor of the compact object and to investigate possible connection to the previous studied case of Centaurus X-4 \citep{gon05a}.

\section{Observations}

\subsection{Secondary spectrum}

As reported in \citet{cas10}, we obtained ten 1800-3600s spectra with the Utrecht Echelle Spectrograph (UES) attached to the 4.2-m William Herschel Telescope (WHT) at the Observatorio del Roque de Los Muchachos on 1999 July 25-26, covering the spectral regions $\lambda\lambda 5300-9000$\AA \space at a resolving power $\lambda/\Delta\lambda \approx 36,000$. 
All the images were processed following standard debiasing and flat-fielding, and the spectra subsequently extracted using conventional optimal extraction techniques in order to optimise the signal-to-noise ratio of the output. For more information about the observations see \citet{cas10}.

The individual spectra were corrected for their radial velocity and combined in order to improve the signal-to-noise ratio. After binning in wavelength in steps of 0.1 \AA, the final spectrum had $S/N\sim70$ in the continuum.

\section{Chemical Analysis}
\subsection{Stellar parameters}

The chemical analysis of secondary stars in LMXB systems is influenced by three important factors: veiling from the accretion disk, rotational broadening and S/N. To obtain the stellar parameters of the secondary star, several moderately strong and unblended absorption lines of FeI and FeII were identified. We selected six absorption features containing relatively strong Fe lines with different excitation potentials. In order to compute synthetic spectra for these features, we adopted the atomic line data from the Vienna Atomic Line Database~\citep[VALD][]{pis95} and used a grid of local thermodynamic equilibrium (LTE) models of atmospheres \citep{kur93}. Synthetic spectra were then computed using the LTE code MOOG 2013 \citep{sne73}. To minimise the effects associated with the errors in the transition probabilities of atomic lines, we adjusted the oscillator strengths, the $\log gf$ values of the selected lines, until we succeeded in reproducing the solar atlas \citep{kur84} with solar abundances \citep{gre96}.

We generated a grid of synthetic spectra for these features in terms of five free parameters, three regarding the star atmospheric model (effective temperature, $T_{\rm eff}$, surface gravity, $\log g$, and metallicity, [Fe/H]) and two more parameters to take into account any possible veiling (the effect of the accretion disk emission in the stellar spectrum). The veiling was defined as a linear function of wavelength and thus described with two parameters, the veiling at 4500\AA, $f_{4500}=F_{\rm disk}/F_{\rm cont, star}$, and the slope, $m_{0}$.
These parameters were modified according to the steps and ranges given in Table~\ref{tabpar}. A rotational broadening of $34.6$ $\rm km$ $\rm s^{-1}$ and a limb darkening of $\epsilon=0.5$ were assumed based on \citet{cas10}. A fixed value of $\xi=2$ $\rm km$ $\rm s^{-1}$ for microturbulence was adopted.

These parameters were modified according to the steps and ranges given in Table~\ref{tabpar}. A rotational broadening of $34.6$ $km$ $s^{-1}$ and a limb darkening of $\epsilon=0.5$ were assumed based on \citet{cas10}. A fixed value of $\xi=2$ $km$ $s^{-1}$ for microturbulence was adopted.
\bigskip
\begin{table}
\begin{center}
\caption{Ranges and steps used to obtain stellar parameters.\label{tabpar}}
\begin{tabular}[c c c ]{| c  c  c |}
\hline
\textbf{Parameter}	&	\textbf{Range}	&	\textbf{Step} 	\\  \hline	 	\hline 
$T_{\rm eff}$ 	&	$5500 - 7500$~K	&	100 K 	\\  
$\log g$ & $1.5 - 4.0$ & 0.1 \\ 
${\rm [Fe/H]}$ & $0.0 - 1.0$ & 0.1 \\ 
$f_{4500}$ & $0.0 - 2.0$ & 0.1 \\ 
\textbf{$m_{0}$} & $0.00000 - 0.00042$ & 0.00004 \\ \hline
\end{tabular}
\medskip
\end{center}
\end{table}

\begin{figure*}
\centering
\includegraphics[scale=1.00]{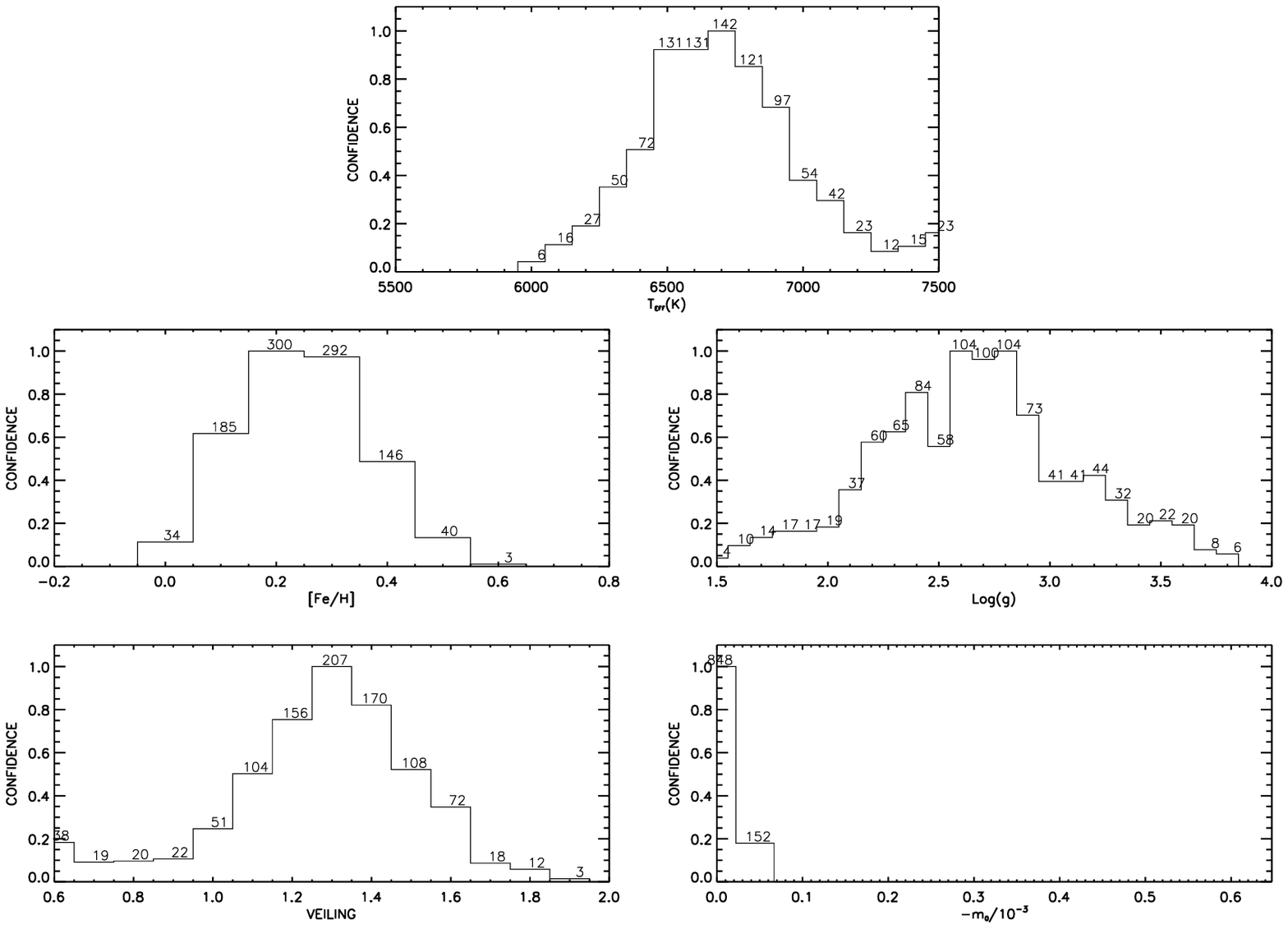}
\caption{Distribution obtained for each parameter using Monte Carlo simulations with five free parameters. The bottom-right panel shows the distribution obtained for the veiling slope $m_{0}$, given as $-m_{o}/10^{-3}$ in units of \AA. The labels at the top of each bin indicate the number of simulations consistent with the bin value. The total number of simulations was 1000.\label{figpar5}}
\end{figure*}

\begin{figure*}
\centering
\includegraphics[scale=1.00]{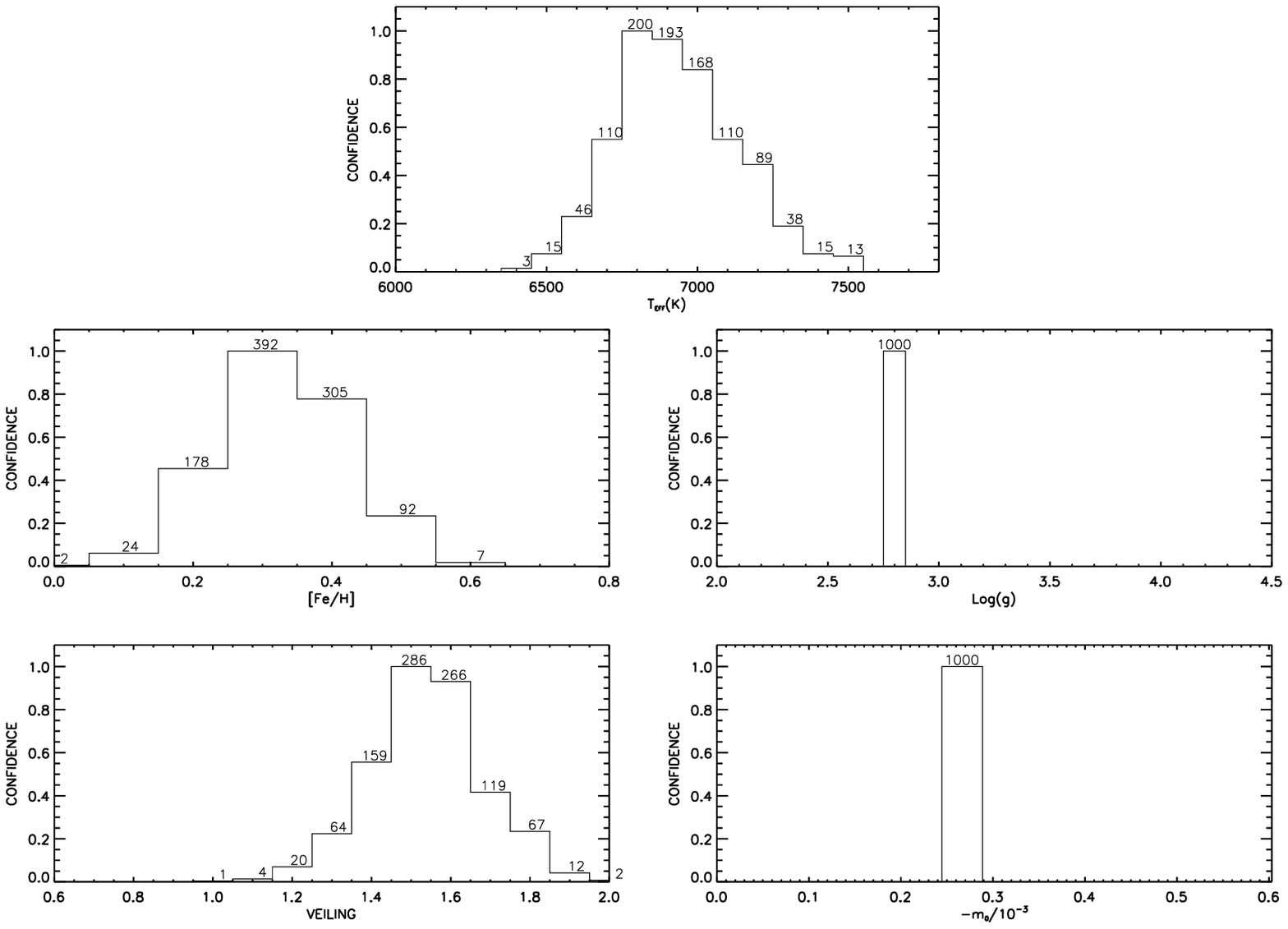}
\caption{Distribution obtained for each parameter using Monte Carlo simulations with four free parameters. The bottom-right panel shows the distribution obtained for the veiling slope $m_{0}$, given as $-m_{o}/10^{-3}$ in units of~{\AA}. The labels at the top of each bin indicate the number of simulations consistent with the bin value. The total number of simulations was 1000.\label{figpar4}}
\end{figure*}

The observed spectrum was compared with the 750,000 synthetic spectra in the grid via a $\chi^{2}$ minimisation procedure that provided the best model fit. Using a bootstrap Monte Carlo method, we define the 1$\sigma$ confidence regions for the five parameters. Confidence regions were determined using 1000 realisations. 
The histograms for the results are shown in Fig.~\ref{figpar5}, where we can see, in particular, a big uncertainty on the value of $\log g$, that affects the rest of our values. 
Given this uncertainty on $\log g$, we decided then to fix this parameter, leaving only four free parameters ($T_{\rm eff}$, [Fe/H], veiling and the slope $m_{0}$) and re-calculate our $\chi^{2}$ minimisation procedure for fixed values of $\log g$, from 1.5 to 4.0. 
We adopt the $\log g$ value that provided us with the lowest uncertainty, i.e. the narrow distributions of values, for the other four free parameters, $\log g=2.8$ dex (see Fig.~\ref{figpar4}). Thus, the most likely values are: $T_{\rm eff}=6900 \pm 200 K $, $\log g = 2.80 \pm 0.20 $, [Fe/H]$=0.35 \pm 0.10$, $f_{4500}= 1.55 \pm 0.15$, and $m_{o}= -0.00027 \pm 0.00004$. 
We decided to adopt an uncertainty on $\log g$ of 0.20 dex based on the shape and width of the histograms of the four free parameters at different $\log g$ values.
These parameters agree with the spectral type and luminosity class of the secondary star reported in the literature~\citep[see e.g.][]{cas10}.
We note the high veiling{\footnote{Note that the veiling defined as $V_\lambda=F_{\rm disk}/F_{\rm total}$ relates with $f_{\lambda}=F_{\rm disk}/F_{\rm cont, star}$ as $V_{\lambda}=f_{\lambda}/(1+f_{\lambda})$.}}  of the accretion disk ($F_{\rm disk}/F_{\rm total} \sim 60 \%$ at 5000~{\AA}), similar to that found in the NSXB Cen X-4~\citep{gon05a}. On the contrary, BHXBs typically show lower disk veilings at less than $\sim 15 \%$ at 5000~{\AA}~\citep{gon11}, except for the BHXB XTE J1118+480 which shows a disk veiling of $\sim 40 \%$ at 5000~{\AA} \citep{gon08b}.

\subsection{Stellar abundances}

Using the derived stellar parameters, we analysed several spectral regions for elements such as Al, Si, S, O, among others. Each of these spectral regions were compared with two templates: WASP-17, with $T_{\rm eff}=6650\pm80$~K, $\log g = 4.45\pm0.15$, and [Fe/H]$=-0.19\pm0.09$, taken from~\citet{tri10}, and HR6189, with $T_{\rm eff}=6216$~K, $\log g = 4.18$, [Fe/H]$=-0.58$, taken from \citet{adi12}, properly broadened with the same rotational broadening as the secondary star in our system. We determined the abundances of those elements by comparing the observed spectrum with a grid of synthetic spectra through a $\chi^{2}$ minimisation procedure. We modified the element abundances, while the stellar parameters and the veiling value were fixed \citep[see e.g.][for further details]{gon11}.

A preliminary estimation of Fe abundance was obtained in the procedure described above. We then performed a more detailed analysis of the abundance of Fe, now taking into account 9 additional Fe lines. 
We obtained Fe abundance of [Fe/H]$=0.27 \pm 0.19$, where the error takes into account the dispersion of the abundances inferred from these features. This error also contains the uncertainty due to the continuum noise. A new model atmosphere was generated with the new Fe abundance in order to perform a detailed spectral synthesis for all the elements under consideration.

Abundances for all the elements are listed in Table~\ref{tababu} and are referred to the solar values adopted from \citet{gre96}, except oxygen, taken from \citet{ecu06}. 
Errors regarding uncertainties in metallicity, $\Delta_{\rm met}$, continuum noise, $\Delta_{\rm cont}$, effective temperature, $\Delta_{T_{\rm eff}}$, gravity, $\Delta_{\log g}$, veiling, $\Delta_{\rm veil}$, and microturbulence, $\Delta_{\xi}$, are listed as well as the error that comes from the dispersion of the abundance measurements from several lines, $\Delta_\sigma=\sigma/\sqrt(N)$, with $N$ equal to the number of lines. All the individual uncertainties except for $\Delta_\sigma$ were determined in the same way as, for instance, in the $T_{\rm eff}$ case: 
$\Delta_{T_{\rm eff}} = (\sum\limits_{i=1}^{N} \Delta_{T_{{\rm eff},i}})/N$. 
The total uncertainty given in Table~\ref{tababu} was derived using the following expression: \\
\\
$\Delta {\rm [X/H]} = (\Delta_\sigma^2+\Delta_{T_{\rm eff}}^2+\Delta_{\log g}^2+\Delta_{\xi}^2+\Delta_{\rm met}^2+\Delta_{\rm cont}^2+\Delta_{\rm veil}^2)^{1/2}$
\\ \\
We stress that the major source for these errors is the inaccuracy in the location of the continuum caused by the S/N and the rotational broadening of the lines. However, we evaluate this source of uncertainty by changing the continuum location by an amount equal to 1/(S/N)~$\sim0.14$.
When the number of lines is high, $\Delta_\sigma$ should contain the uncertainty in the location of the continuum, $\Delta_{\rm cont}$, but due to the quality of the data, the stellar parameters of the secondary star and its rotational broadening, we typically find very few lines able to provide reliable element abundances, except for Fe. 
Most of the lines used in the abundance analysis are weakly blended with lines of other elements (mostly Fe lines). However, the contribution of these blends is typically not dominant.

\begin{figure}
\begin{center}
\includegraphics[scale=0.50]{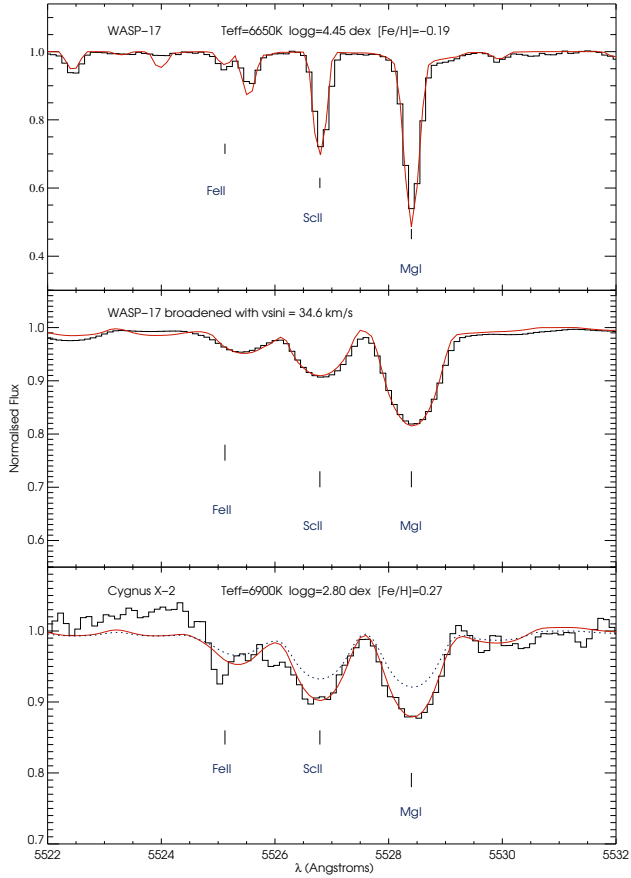} 
\caption{Best synthetic fit to the UES spectrum of the secondary star in the NSXB Cygnus X-2 (bottom panel) and best fit to our template, with and without rotational broadening (middle and top panels). Synthetic spectra is computed for best-fit abundances (solid line) and for solar abundances (dashed line).\label{figmg}}
\end{center}
\end{figure}

\begin{figure}
\begin{center}
\includegraphics[scale=0.50]{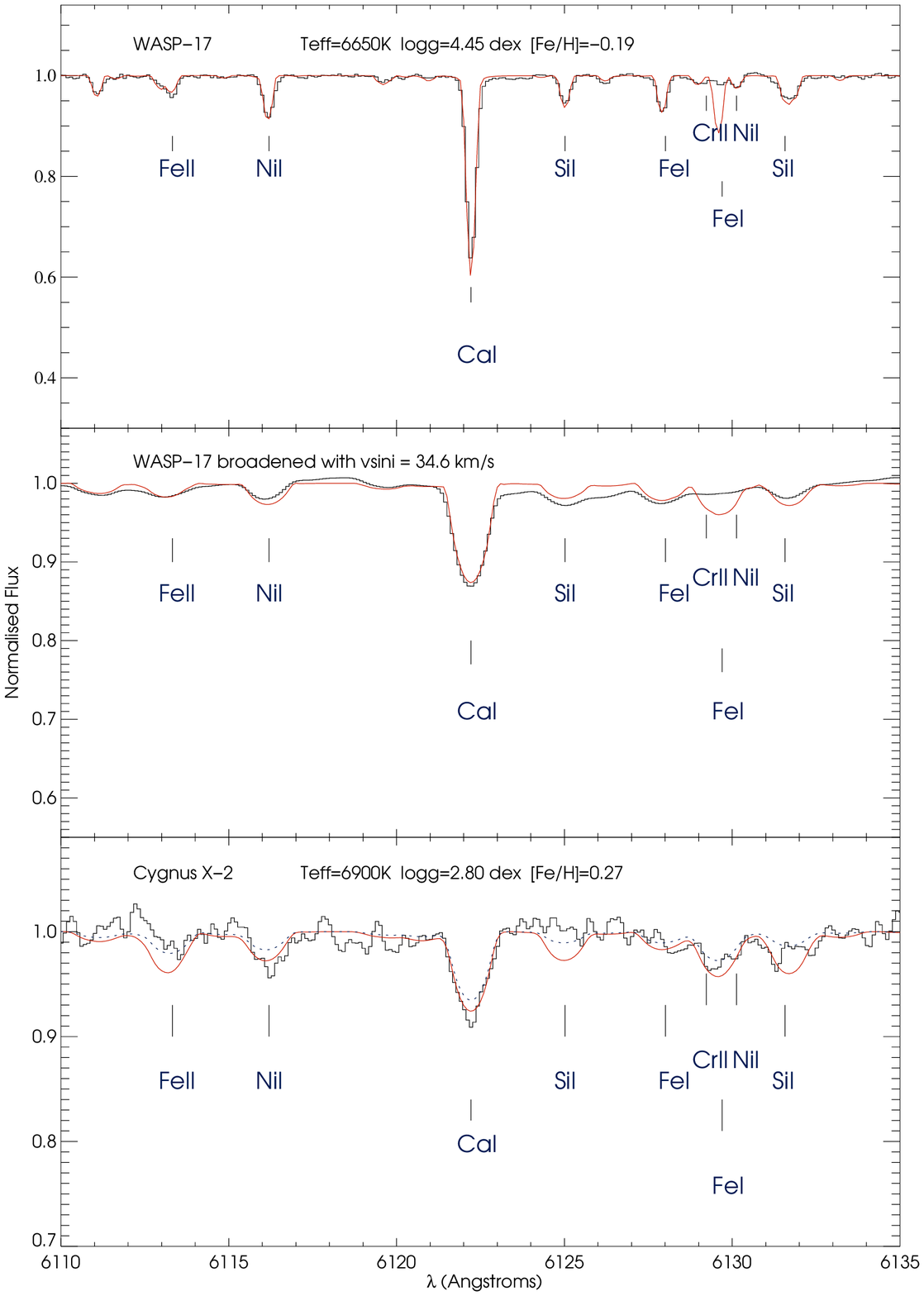} 
\caption{The same as Fig.~\ref{figmg} but for another spectral range.\label{figcani}}
\end{center}
\end{figure}

\begin{figure}
\begin{center}
\includegraphics[scale=0.50]{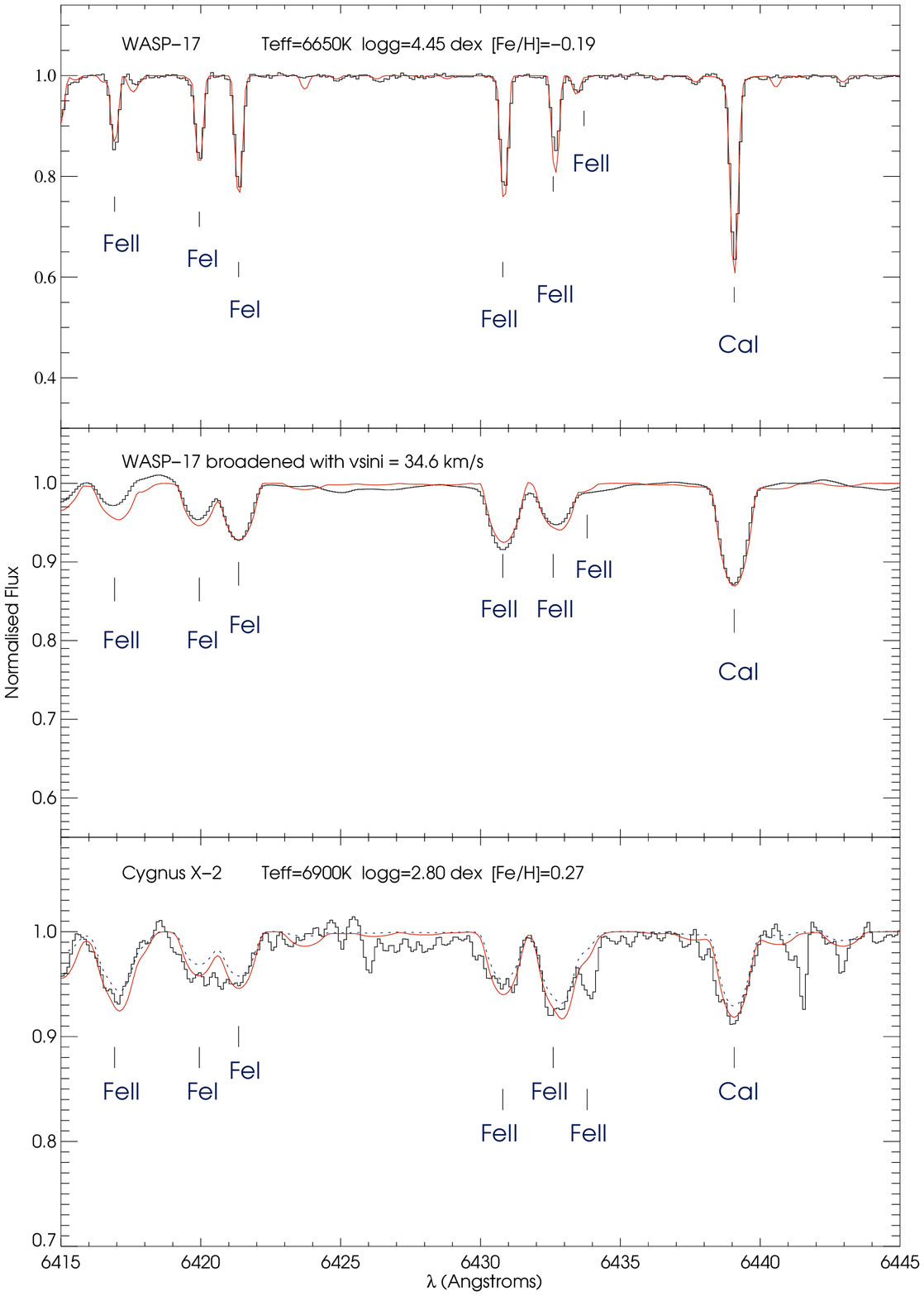} 
\caption{The same as Fig.~\ref{figmg} but for another spectral range.\label{figcafe}}
\end{center}
\end{figure}

\begin{figure}
\begin{center}
\includegraphics[scale=0.50]{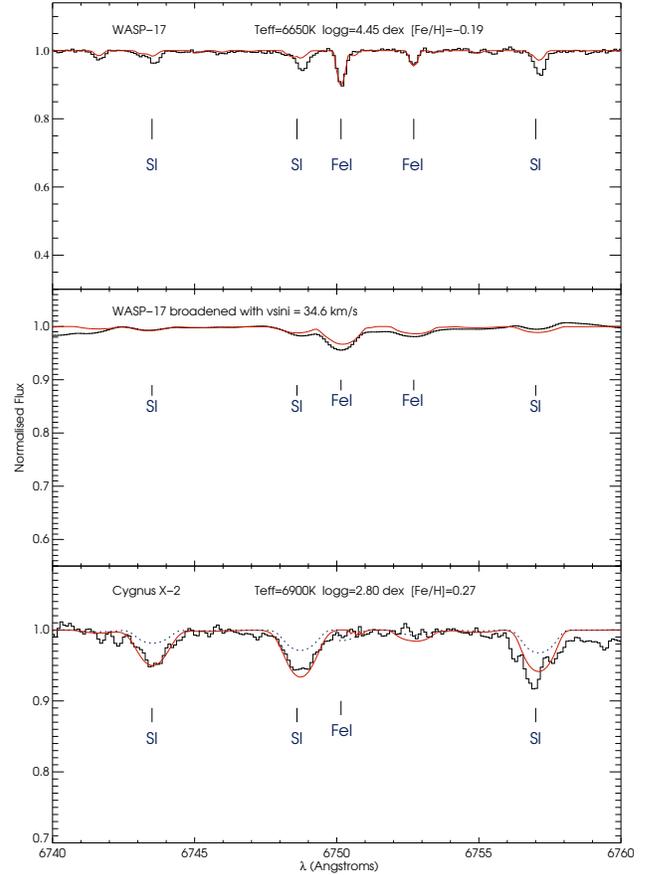} 
\caption{The same as Fig.~\ref{figmg} but for another spectral region,
showing the three spectral features of sulphur.\label{figsu}}
\end{center}
\end{figure}

\begin{figure}
\begin{center}
\includegraphics[scale=0.50]{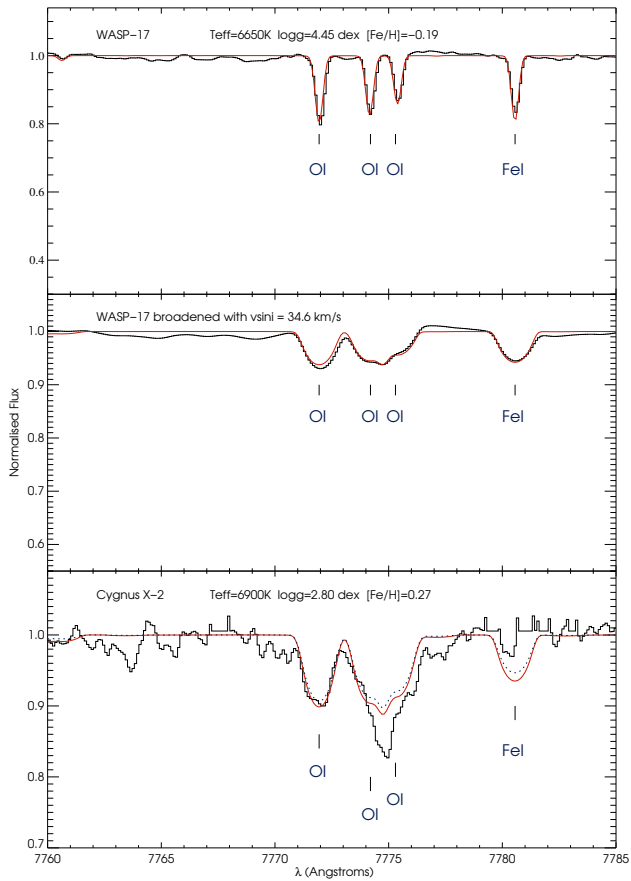} 
\caption{The same as Fig.~\ref{figmg} but for another spectral region
showing the near-IR OI triplet.\label{figox}}
\end{center}
\end{figure}

\begin{figure}
\begin{center}
\includegraphics[scale=0.50]{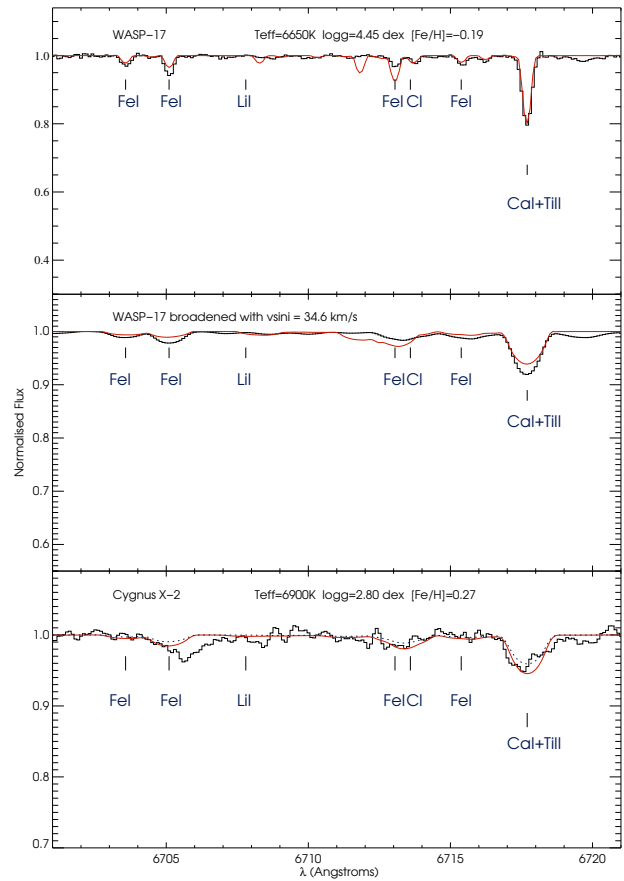} 
\caption{The same as Fig.~\ref{figmg} but for another spectral region
showing the LiI 6708\AA \space feature.\label{figli}}
\end{center}
\end{figure}

In Figs.~\ref{figmg}~$-$~\ref{figli}, we show the best fit model synthesis in comparison with the observed spectra. We also depict, for comparison, the same spectral region of one of the templates,  properly broadened using the rotational profile of our star, together with the best fit synthetic spectra which typically show abundances close to solar values.

\begin{table*}
\caption{Chemical abundances and uncertainties due to the errors of $\Delta_{\rm met}=0.05$~dex, $\Delta_{\rm Teff}=200$~K, $\Delta_{\rm log g}= 0.20$~dex, $\Delta_{\rm \xi}= 0.50$~km~s$^{-1}$ and $\Delta_{\rm vel} = 0.15$. The oxygen abundance has been corrected for NLTE effects, $\Delta_{\rm NLTE}=-0.3$ dex based on the NLTE corrections in \citet{prz00}. Solar abundances were adopted from \citet{gre96} for all elements except oxygen, which was taken from \citet{ecu06}. Li abundance is expressed as: $\log \epsilon({\rm Li})_{\rm NLTE} = log[N({\rm Li})/N({\rm H})]_{\rm NLTE}+12$.\label{tababu}}
\centering
\scalebox{0.8}[0.8]{
\begin{tabular}{| l  c  c  c  c  c  c  c  c  c  c  c  c  c  |}
	\hline 
\textbf{ Element }& \textbf{$[\rm X/\rm H]$} &\textbf{$[\rm X/\rm Fe]$} & \textbf{$\sigma$} &\textbf{$\Delta_{\rm met}$} &\textbf{$\Delta_{\rm cont}$} & \textbf{$\Delta\sigma$} &\textbf{$\Delta_{\rm Teff}$} & \textbf{$\Delta_{\rm log g}$} & \textbf{$\Delta_{\rm \xi}$} &\textbf{$\Delta_{\rm vel}$} & \textbf{$\Delta [\rm X/\rm H] $} & \textbf{$\Delta [\rm X/\rm Fe]$} & \textbf{$\rm n$}\\
	\hline \hline
Fe & 0.27 & ... & 0.19 &-0.06 & 0.17 & 0.05 & 0.05 & 0.04 & 0.00 & 0.08 & 0.19 & ... &15 \\
O & 0.07 & -0.20 & 0.00 &0.05 & 0.30 & 0.00 & 0.10 & 0.10 & 0.00 & 0.10 & 0.35 &0.20 & 1 \\
Mg & 0.87 & 0.60 &0.00&0.01&0.10&0.00&0.08&0.01&0.01&0.20&0.24&0.16& 1 \\
Al & 0.42 &0.15 &0.05&0.08& 0.20 & 0.04 & 0.08 & 0.02 & 0.00 &0.08 & 0.24 & 0.15 & 2 \\
Si & 0.52 & 0.25 & 0.05&0.00& 0.19 & 0.03 & 0.06 & 0.00 & 0.00 &0.08 & 0.22 &0.07 & 3 \\
S & 0.52 & 0.25 &0.08&0.01& 0.20 & 0.05 & 0.05 & 0.03 & -0.01 &0.12 & 0.24 &0.09 & 3 \\
Ca & 0.27 & 0.00 & 0.05 & 0.01 & 0.30 & 0.03 & 0.10 & 0.02 &-0.01 & 0.10 & 0.33 & 0.16 & 4 \\
Ti & 0.59 &0.32 &0.04&0.01& 0.28 & 0.03 & 0.05 & 0.08 & 0.00 & 0.10 & 0.31 & 0.14 & 2 \\
Sc & 0.42 & 0.15 &0.00 &0.01&0.20&0.00&0.05&0.07&-0.04&0.20&0.30&0.15& 1 \\
Ni & 0.52 & 0.25 &0.00 &0.01&0.20&0.00&0.09 &0.01&-0.01&0.15&0.27&0.10& 1 \\
Li 	&	$<1.48$	& $<1.21$& - &-&-&-&-&-&-&-&-&-& 1 	\\		
	\hline 
\end{tabular}}
\medskip
\end{table*}

\subsection{Lithium}
Lithium abundance was derived from the resonance line at $6708$ \AA (see Fig.~\ref{figli}). In our spectra, this line is barely detected, giving an upper-limit in the lithium abundance of $\log\epsilon({\rm Li})_{\rm NLTE} < 1.48$. The effective temperature of this star implies that is near the limits of the \textit{lithium gap} \citep[between 6400-6800K][]{boe86}, so Li depletion may have occurred due to diffusion. According to models by \citet{mic86}, that explain the Li gap in the Hyades~\citep{boe86} (with an age of $\sim600$~Myr), the depth of the convective zone of stars in the $T_{\rm eff}$ range 6400-7000~K increases rapidly as $T_{\rm eff}$ decreases. The radiative acceleration is smaller at deeper layers of the star, and this makes more efficient Li diffusion at 6500K than at 7000K. The $T_{\rm eff}$ determination of the secondary star in Cygnus X-2 is $6900\pm200$~K, and therefore perfectly consistent with the Li gap, providing an explanation for the low Li abundance regardless any consideration or assumption on the age of the system. 

On the other hand, the NSXB Centaurus X-4, with a K-type secondary star of $T_{\rm eff}=4500\pm100$~K, shows an unexpectedly high lithium abundance of $\rm log\epsilon({\rm Li})_{\rm LTE}=3.06$, that could be explained if the Cen X-4 system is younger than the Pleiades cluster, whose age is $\sim$120 Myr \citep{gon05a,gon05b}.
Possible Li production mechanism has been explored through the measurement of $^6{\rm Li}/^7{\rm Li}$ isotopic ratio of the secondary star in Cen X-4, which mostly supports Li preservation scenarios based on some mechanisms which inhibit Li destruction due to the fast rotation of the secondary star~\citep{cas07}. The different spectral types of secondary stars in Cen X-4 and Cyg X-2 makes it difficult to compare their Li abundance and evolutionary scenario.
The BHXB Nova Scorpii 1994 has a more similar secondary star, an F-type star with $T_{\rm eff}=6100\pm200$~K, although it has a relatively high lithium abundance of $\rm log\epsilon({\rm Li})_{\rm LTE}=2.16$~\citep{gon08a}. However, this Li abundance is consistent with that of main-sequence F-type disk stars with ages in the range $\sim1-3$~Gyr and similar metallicity~\citep[e.g.][]{boe87}.

\subsection{Oxygen\label{secoxy}} 
Oxygen abundance was derived from one component of the OI near-infrared triplet, OI~7771~{\AA} (see Fig.~\ref{figox}), which is certainly well reproduced with the synthetic spectra. Other components of the OI triplet could not be used because of the limited signal-to-noise and possibly the inaccurate sky subtraction. We note that the atomic data for O I lines were adopted from~\citet{ecu06}. These authors slightly modified the oscillator strengths in order to obtain a solar oxygen abundance in LTE of $\log \epsilon ({\rm O})_{\odot} = 8.74$. The best fit in LTE of the OI~7771~{\AA} line in the secondary star of Cygnus X-2 gives an abundance of [O/H]$_{\rm LTE}=0.37\pm0.35$.
This value provides an abundance ratio $[\rm O/\rm Fe]_{\rm LTE} =0.10 \pm 0.20$, roughly consistent with the solar value. The uncertainty due to continuum location, $\Delta_{\rm cont}=0.30$, is quite high. The O I $\lambda$7771{\AA} triplet suffers from appreciable non-LTE (NLTE) effects~\citep[see e.g.,][]{ecu06}. For the stellar parameters and metallicity of the secondary star in Cygnus X-2, the NLTE corrections\footnote{$\Delta_{\rm NLTE} = \log \epsilon ({\rm X})_{\rm NLTE} - \log \epsilon ({\rm X})_{\rm LTE}$} are expected to be as strong as $\Delta_{\rm NLTE}\sim-1.0$~dex according to their NLTE computations~\citep[see Fig. 1 in][]{ecu06}. These authors do not include inelastic collisions with hydrogen atoms which tend to decrease the NLTE corrections.  \citet{tak03} evaluate the impact of these collisions on the strength of non-LTE effects in the OI triplet. They model this aspect following the prescription in \citet{kis93}, who make use of the generalisation in \citet{ste84} of the Drawin's formula to account for the inelastic collisions with hydrogen atoms~\citep{dra69}. According to the NLTE computations in \citet{tak03}, the expected NLTE corrections for the stellar parameters and metallicity of the secondary star in Cyg X-2 would be $\Delta_{\rm NLTE}\sim\;-0.8\;-\;-1.0$~dex from their Fig.~11, even including hydrogen collisions scaled by a factor of 1/3 (i.e. $h=-0.48$ according to their definition). This factor, $S_H=1/3$, is the preferred value used by \citet{caf08} to estimate the solar oxygen abundance from the near-IR OI triplet, which provides NLTE corrections of $\Delta_{\rm NLTE}\sim-0.22$~dex for the solar OI lines. This value is very similar to those derived in \citet{kis93} for the Sun. We have not found published values of NLTE corrections for oxygen in late A-, early F-type stars, and the figures given in \citet{tak03} and \citet{ecu06} do not include stars hotter than 6500~K.

We note that applying the $\Delta_{\rm NLTE}\sim-1.0$~dex to the LTE oxygen abundance would lead to [O/H]$_{\rm NLTE}\sim-0.63$, which implies that oxygen is significantly depleted in this star. Thus, this star would have an anomalously low abundance ratio [O/Fe]~$\sim -0.9$~dex, and therefore not consistent with the Galactic trend of oxygen as measured in F-, G-, and K-type stars (see Fig.~\ref{figgal}). 
\citet{rob90} found undersolar abundances at about $-0.20\pm0.15$~dex in field stars with $T_{\rm eff}\sim 6700-7000$~K whereas \citet{gar93} analysed stars from the Hyades and Ursa Major clusters and found O abundances slightly enhanced with respect to the solar value by only about $+0.05\pm0.20$~dex in the same $T_{\rm eff}$ regime. \citet{gar93} computed NLTE corrections of about $-0.4 \; - \; -0.6$~dex in this $T_{\rm eff}$ range, depending on oxygen abundance and $T_{\rm eff}$ values, although the strengths of their NLTE effects may be also related to the high solar and stellar oxygen abundances these authors measured. 

\citet{prz00} evaluate NLTE effects on oxygen in BA-type stars with $T_{\rm eff}$ in the range 7500--15000~K. We extrapolate the NLTE values for oxygen triplet given in their figure 5 to the stellar parameters of the secondary star ($T_{\rm eff}\sim6900$~K and $logg\sim2.8$~dex), thus providing $\Delta_{\rm NLTE}\sim-0.30$~dex for oxygen in the secondary star.  The resulting oxygen abundance is [O/H]$_{\rm NLTE}\sim +0.07$ and an abundance ratio [O/Fe]~$\sim -0.20$~dex (see Table~\ref{tababu}). The adopted NLTE correction makes the oxygen abundance of the secondary star roughly consistent with the lower edge of Galactic trend of F-, G-, and K-type stars (see Fig.~\ref{figgal}). However, we would like to stress that the NLTE correction may be larger and this value may be considered as an upper limit to the NLTE effects on the oxygen triplet in the secondary star.

The low oxygen abundance in the secondary star may be explained by diffusion~\citep{mic87}. These authors modelled atomic diffusion in early A-type stars hotter than our case; they claim that diffusion lead to normal CNO abundances, being downward diffusion velocity of heavy elements smaller than that of He (e.g. 30 times smaller for C than for He). 
\citet{gon95} modelled radiative accelerations of CNO atoms in A-, F-type stellar envelopes, to explain CNO abundance anomalies. These authors found that radiative accelerations of CNO atoms are typically smaller than local gravity at the base of the convective zone for a large range of stellar abundances. However, these models produced too depleted oxygen abundances by $1-2$~dex with respect the abundance measurements in \citet{gar93}, and may only offer and explanation for C abundances in stars hotter than 7500~K.
More recently, \citet{ric01} have performed computations including radiative accelerations, atomic diffusion and several levels of turbulence, from light elements as H, He and Li, up to heavy elements as Ti, Fe and Ni. Their Fig.~\ref{figgal} shows the stellar abundance predictions for 1.7~$M_\odot$~star. O and S appear severely depleted from -0.5 dex down to -3~dex, depending of the level of turbulence, being more depleted as the turbulence decreases. Similarly, Mg, Si and Ca follow this behaviour although they stay within the range -0.3 and -1.0 dex with respect to solar values. Al is not so dependent on turbulence and it typically shows solar or slightly undersolar abundances. Finally, Ti and Fe-peak elements tend to show the opposite behaviour, revealing enhanced abundances from solar up to +1.5 dex. 

The abundance pattern of the secondary star does not seem to be explained by any of these models, as the secondary star shows large enhancements in Mg, Si, S, and Ti but moderate enhancement in Ni. Low O abundance and high Ti abundance could be explained by these models, but we think the ejection of nucleosynthetic products during SN explosion that originated the neutron star could be the responsible for the abundance pattern of the secondary star in Cygnus X-2 (see Section 4).

On the other hand, \citet{pod00} have suggested an evolutionary scenario in which the initial mass of the secondary star could be as high as 3.5~$M_{\odot}$. They argue that after the SN explosion, the subsequent evolution is driven by strong mass loss. Thus, after the secondary star recovers thermal equilibrium, its surface may be exposed to material that has undergone partial CNO burning, and its composition should show the signature of CNO processing (enhanced nitrogen, decreased carbon and oxygen). This may explain the low O abundance seen in this system, but unfortunately we do not have clean available N and C features to test this model. 
However, the O abundance is not expected to decrease too much compared with Carbon~\citep{cla83}. Thus, the CNO-processed material should mostly be C under-abundant and N over-abundant.
We note that a similar scenario has been proposed for the BHXB XTE J1118+480 to try to explain the weak Carbon emission lines coming from the accretion disk seen in UV spectra~\citep{has02}. Unfortunately, neither O nor C or N, have been measured in the secondary star of XTE J1118+480 yet~\citep{gon08b}.

\subsection{Other $\alpha$-elements}
Magnesium abundance was derived from the Mg I feature at 5528~{\AA} (see Fig.~\ref{figmg}). We note that the feature is very sensitive to stellar parameters. As in the case of oxygen, the S/N plays an important role and makes it difficult to correctly place the continuum.  \citet{zha98,zha00} evaluate the NLTE effects of this Mg line and found always positive but small NLTE corrections of about $\Delta_{\rm NLTE}\sim+0.05$~dex that we do not take into account.
Sulphur was measured from the SI features at $6743-57$~{\AA} that are well detected at this S/N (see Fig.~\ref{figsu}). Silicon, calcium and titanium were measured from several lines as well. In Figs.~\ref{figcani} and~\ref{figcafe} we depict several spectral regions with some Ca and Si features, together with FeI and FeII features.

\subsection{The odd-Z elements: Al, Sc}
Aluminium abundance is derived from two features whereas scandium is measured from one single line displayed in Fig.~\ref{figmg}. The Sc abundance may be affected by the location of the continuum which makes it difficult to fit at the S/N level of the observed spectrum.

\subsection{Fe-peak elements}
Iron was derived from 15 features which makes quite reliable the abundance measurement. The Ni abundance, however, was measured from just one single although isolated and well detected line (see Fig.~\ref{figcani}).

\section{Discussion\label{secdis}}
The metallicity obtained for this object is higher than solar, and similar to the metallicity obtained in another neutron star X-ray binary Cen X-4 \citep[see][]{gon05a}.
Although may not be statistically significant, the metallicities of BHXBs seems to be lower, with a weighted mean [Fe/H] of $\sim 0.09\pm0.06$  ($\sigma_N=0.06$, for $N=5$~BHXBs), than that of NSXBs, with a weighted mean [Fe/H] of $\sim 0.24\pm0.09$  ($\sigma_N=0.02$, for $N=2$~NSXBs). This may indicate that indeed a relatively high amount of Fe is always ejected during supernova explosions that form neutron stars whereas for black holes, the initial mass of the collapsing material, the so-called mass cut ($m_{\rm cut}$), may be high enough to prevent large amounts of Fe to escape from the collapsing material and/or to avoid possible and efficient mixing processes between the fallback matter onto the compact object and the ejected matter. 
In this sense, the detection of high Fe abundances in secondary stars of NSXBs may be an indication of an explosive event.

\begin{figure*}
\begin{center}
\includegraphics[scale=0.7, angle=90]{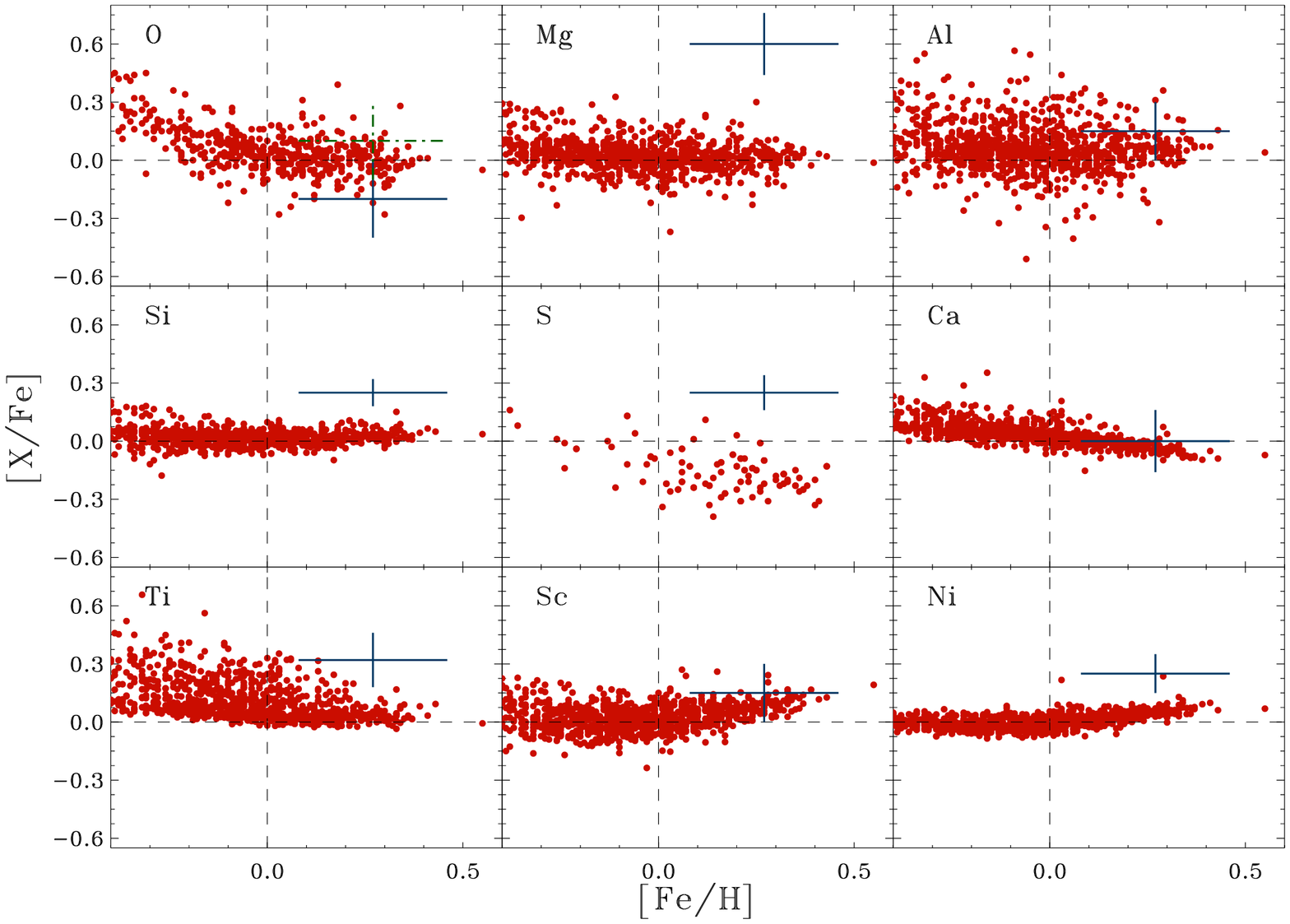} 
\caption{Element abundance ratios [X/Fe] of the secondary star in the NSXB Cygnus X-2. Crosses indicate our obtained abundance values with errors, while full dots are the abundances of galactic field stars from \citet{adi12}, except for oxygen and sulphur whose data were taken from \citet{ber14} and \citet{ecu04}, respectively.For oxygen we display two values of the [O/Fe] abundance ratio, one in LTE (dotted-dashed cross) and the other one in NLTE (solid cross).\label{figgal}}
\end{center}
\end{figure*}

We search for anomalies in the abundance pattern of the secondary star by comparing our results with Galactic trends (see Fig.~\ref{figgal}), adopted from \citet{adi12}. For oxygen, we used the latest results from \citet{ber14}. We display the [O/Fe] abundance ratio in LTE (dotted-dashed cross) and NLTE (solid cross) (See Section 3.4 for further details).  In the following sections we will be only using or refering the derived [O/Fe] abundance ratio in NLTE. Results for sulphur were taken from \citet{ecu04}. As it is shown in Fig.~\ref{figgal}, most of the elements in Cygnus X-2 show over-abundances when compared with Galactic trends, with the exception aluminium, aalcium and acandium, which are consistent with those trends. In Table~\ref{tabgal} we depict the element abundance ratios, [X/Fe], of the secondary star in comparison with the average values of F-, G-, and K-type stars belonging to Galactic thin disk in the relevant range of metallicities ($0.16 < $[Fe/H]$ < 0.38$). The errors on abundances ratios are typically smaller than absolute errors since different elements show similar sensitivities to changes for instance in the stellar parameters (see 12th and 13th columns of Table~\ref{tababu}, and third column of Table~\ref{tabgal}). Table~\ref{tabgal} shows clear enhancements in Mg, Si, S and Ni at more than 3$\sigma$, Ti at about 2$\sigma$. Al, Ca, and Sc are consistent at 1$\sigma$ with the Galactic trends and finally O is under-abundant but consistent at 1$\sigma$. This abundance pattern cannot be explained using models with atomic diffusion, radiative acceleration and turbulence as seen in Section~\ref{secoxy}. In the following sections we will try to explain these abundances in the context of the supernova explosion that happened very early in the evolutionary scenario of the NSXB Cygnus X-2.

\begin{table}
\caption{Element Abundance Ratios in Cyg X-2.\label{tabgal}}
\raggedright
\scalebox{0.75}[0.8]
{
\begin{tabular}{| c  c  c  c  c  c |}
	\hline 
\textbf{ Element }& \textbf{$[\rm X/\rm Fe]_{\rm Cyg X-2}$}
&\textbf{$ \Delta_{[\rm X /\rm Fe], \rm Cyg X-2}$} &
\textbf{$[\rm X/\rm Fe]_{\rm stars}$} &\textbf{ $\sigma_{\rm
met}$} &\textbf{$\Delta_{\rm \sigma,\rm stars}$} \\
	\hline \hline
O & -0.20 & 0.20&      0.01  &      0.09 &      0.01 \\
Mg & 0.60 & 0.16&        0.01  &      0.07 &      0.00 \\
Al & 0.15 & 0.15 &        0.05  &      0.10 &      0.01 \\
Si & 0.25 & 0.07 &        0.02  &      0.03 &      0.00 \\
S & 0.25 & 0.09&       -0.18  &      0.09 &      0.01 \\
Ca & 0.00 &0.16 &       -0.02  &      0.03 &      0.00 \\
Ti & 0.32 & 0.14 &        0.05  &      0.05 &      0.00 \\
Sc & 0.15 & 0.15 &        0.06  &      0.05 &      0.00 \\
Ni & 0.35 & 0.10 &        0.03  &      0.04 &      0.00 \\
\hline
\end{tabular}}
\medskip
\end{table}

\subsection{Spherical SN explosion}
It has been proposed that LMXBs begin their lives as wide binaries and evolve through a common-envelope phase in which the companion spirals into the massive star's envelope \citep[see e.g.][]{heu83,koo87,nel01}. The helium core of the massive star evolves and ends its live as a supernova, leaving behind a neutron star or a black hole remnant. Part of the ejected mass in the SN explosion may have been captured by the secondary star, as has been found in other LMXBs such as Nova Sco 94~\citep{isr99,gon08a}. We also consider this scenario for Cygnus X-2.

Cygnus X-2 is quite a particular X-ray binary. The stellar parameters obtained suggest that the secondary is a F2IV star. The primary star is a the neutron star of $M_{\rm NS,f}=1.71\pm0.20M_{\odot}$ and the mass ratio of the system is $q_{f}=0.34\pm0.02$ \citep{cas10}, yielding a secondary mass $M_{2,f}=0.60\pm0.13M_{\odot}$. 
A main sequence star of this mass would have an effective temperature of ~4000 K which is in conflict with the observed F-type spectrum \citep{dan94,sie00}. 

The secondary star in Cygnus X-2 might have lost a significant amount of its initial mass through mass-transfer \citep{king99,pod00}.\citet{pod00} propose an evolutionary scenario where the secondary star has an initial mass of $M_{2,i}=3.5M_{\odot}$, and suffers a large mass loss near the end of the main sequence. Thus we adopt different scenarios for the SN model, as the initial mass of the secondary is unknown. The input parameters of our models are basically $M_{2,i}=[2.0 - 3.5]M_{\odot}$, the initial mass of the neutron star is $M_{{\rm NS,}i}\sim1.4 M_{\odot}$ and the radius of the secondary star is obtained using pre-main sequence (PMS) stellar evolutionary tracks from \citet{sie00}. In order to get the secondary stellar radius at the time of the SN explosion we assume that it happened about 7 Myr after the formation of the system~\citep[][and references therein]{gon05a}.
Binary systems like Cygnus X-2 will survive a spherical SN explosion if the ejected mass, $\Delta M$ satisfies this condition: $\Delta M=M_{\rm He}-M_{\rm NS,i} \leq (M_{\rm He}+M_{2,i})/2$. This implies a mass of the He core before the SN explosion of $M_{\rm He}\leq[4.8 - 6.3]M_{\odot}$ for secondary masses $M_{2,i}=[2.0 - 3.5]M_{\odot}$. Using the expressions given by \citet[][and references therein]{por97}, the initial mass of the progenitor star ($M_{1}$) should be $M_{1} \leq [19 - 23] M_{\odot}$ for $M_{2,i}=[2.0 - 3.5]M_{\odot}$, and a radius of $R_{\rm He}  \leq [1.1 - 1.5] R_{\odot}$ for the helium core.

Assuming a pre-SN circular orbit and an instantaneous spherically symmetric ejection (that is, significantly shorter than the orbital period), one can estimate the pre-SN orbital separation, $a_0$, using the relation given by \citet{heu84}: $a_0 = a_{c,i} \mu_f$, where $\mu_f = (M_{{\rm NS,}i} + M_{2,i})/(M_{\rm He} + M_{2,i})$. We find $a_0 \simeq 5.7 - 6.5 R_{\odot}$, for the adopted values of $M_2=[2.0 - 3.5]M_{\odot}$, $M_{\rm He} = 4 M_{\odot}$ and $a_{c,i} =10 R_{\odot}$. 
Secondary stars with masses of $2.0 M_{\odot}$ and $3.5 M_{\odot}$ would be still in their PMS stage having radii of about $\sim 2.4 R_{\odot}$ and $\sim 2.2 R_{\odot}$ respectively. Some of the matter ejected in the SN explosion is expected to be captured by the secondary star, thus polluting its atmosphere with fresh SN nucleosynthetic products. The fraction of the matter ejected in the direction of the secondary star that is finally captured is controlled by the parameter $f_{\rm cap}$. 
The amount of mass deposited on the secondary can be estimated as $m_{\rm cap} = \Delta M (\pi R^2_{2,i} /4 \pi a_0^2) f_{\rm cap}$. The ejected mass, $\Delta M$ is equal to $M_{\rm He} - M_{{\rm NS,}i}$. The matter captured, $m_{\rm cap}$, typically of the order of $\sim 0.03 - 0.10 M_{\odot}$ for $M_{2,i}=2 M_{\odot}$ and $f_{\rm cap}=0.3-0.9$, has larger molecular weight than the pre-explosion content of the secondary star, so it is expected to be quickly mixed with the entire secondary star~\citep[for further details see e.g.][and references therein]{gon11}.

We adopt a current orbital distance of $a_{c,f} \simeq 25R_{\odot}$, as propose by \citet{vrt03}. Considering that the secondary must have experienced mass and angular momentum losses during the binary evolution until it has reached its present configuration, we propose that the post-SN orbital separation after tidal circularization of the orbit should lie at about $10 R_{\odot}$. We note that most of the mass lost from the secondary star, from its initial mass of about $M_{2,i}=[2.0-3.5]M_\odot$ to reach the current estimated mass $M_{2,f}\sim 0.6 M_\odot$, must have been lost from the X-ray binary system. The small difference between the initial, $M_{{\rm NS,}i}\sim 1.4M_{\odot}$, and final mass, $M_{{\rm NS,}f}\sim 1.7M_{\odot}$, of the compact object do not allow high accretion efficiencies during the evolution of the system.
We consider spherically symmetric supernova models for a $\sim 16M_{\odot}$ progenitor star with a $M_{\rm He}\sim4.0 M_{\odot}$ He core~\citep{mae02, gon05a}, and an initial pre-explosion mass for the secondary star of $M_{2,i}=2.0 M_{\odot}$. 
Different values for the initial mass of the secondary star at e.g. about $M_{2,i}=3.5 M_{\odot}$ would provide similar results for lower $a_{c,i}$ and higher $f_{\rm cap}$ values.We do not have any model with masses greater than~$M_{\rm He}=4.0 M_{\odot}$ and smaller than~$M_{\rm He}=16.0 M_{\odot}$, but we think this $\sim 16M_{\odot}$ SN explosion model satisfies our requirements~\citep[see][for further details on these SN models]{gon05a}.

\begin{figure}
\begin{center}
{\includegraphics[scale=0.5]{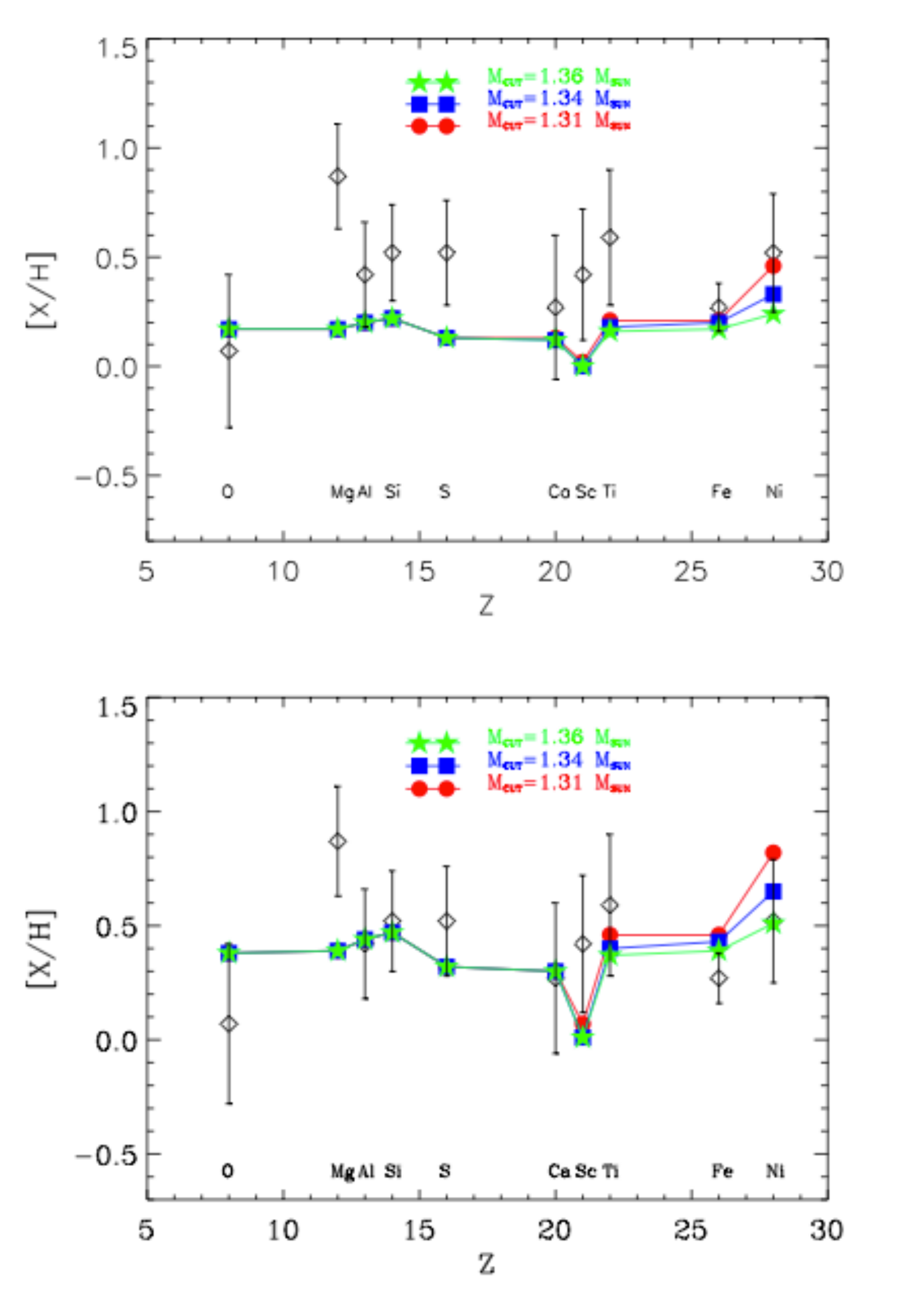}}
\caption{Expected abundances in the secondary atmosphere due to the pollution caused by a spherically symmetric core-collapsed SN explosion model, taking into account $M_{\rm He}=4.0 M_{\odot}$, $M_{2,i}=2.0 M_{\odot}$ and a fixed value of $f_{\rm cap}=0.3$ (top panel) and $f_{\rm cap}=0.9$ (bottom panel). In both cases an orbital distance of $a_{c,i}=10R_{\odot}$ and a kinetic energy of $E_{K} = 1\times10^{51}$erg have been adopted.\label{figssn}}
\end{center}
\end{figure}

One could then obtain different values for the expected abundances by modifying $a_{c,i}$ and $f_{\rm cap}$ in each case. We assume that the secondary star had solar abundances before pollution. The expected abundances for the secondary star after being polluted by supernova products (assuming solar-abundance models) are given in Table~\ref{tabsn}. Best model fits are shown in Fig.~\ref{figssn}, for $f_{\rm cap}=0.3$ in the top panel and $f_{\rm cap}=0.9$ in the bottom panel. To qualify the comparison between observed and modeled abundances we also provide two quality factors, defined as follows:
\begin{equation}
Q = \sum_{i}^{N}([\rm X/H]_{\rm obs,i}-[\rm X/H]_{\rm mod,i})^2]/ \Delta[\rm X/H]_{\rm obs,i}^2) / \nu
\end{equation}

where $N$ is the number of elements: for $Q$ is 10 and for $Q'$ we only considered those elements with enhanced abundances, Mg, Si, S, Ti, Ni and Fe. $\nu$ gives the dregrees of freedom, i.e. $N-n-1$, where $n$ is the number of free model parameters. We adopt as free parameters $f_{\rm cap}$ and $m_{\rm cut}$, and thus $n=2$ .

Our model computations take into account three different mass cut values, $m_{\rm cut}$, which are typically very similar to initial and final mass of the neutron star. As clearly seen in Fig.~10, dependence of the mass cut is only important for elements with Z greater than 20, in particular, Ca, Ti and, preferentially, Fe-peak elements tend to be more abundant for lower $f_{\rm cap}$ values. 
In order to well fit the low oxygen abundance a relatively low capture efficiency, $f_{\rm cap} \sim 30\%$ is required, whereas more of the elements are reasonably reproduced for higher $f_{\rm cap}$ values. In particular, a mass cut  $m_{\rm cut} \sim 1.3 M_{\odot}$ together with a high capture efficiency ($f_{\rm cap} \leq 90\%$) seems to reproduce the observed element abundances except for magnesium which shows too high abundance. The quality factors $Q$ and $Q'$ show that only those spherical models with high capture efficiency and possibly mass cut at about 1.36~$M_{\odot}$ can reproduce the observed abundances but any of them can fit Mg abundance. However, due to the uncertainties in our explosion model we cannot exclude the spherically symmetric supernova scenario.

\begin{table*}
\caption{Expected abundances in the secondary atmosphere due to the pollution caused by the core-collapse SN explosion in three plausible cases: (i) spherically symmetric explosion model ($Spherical$), (ii) non-spherically symmetric explosion ($Aspherical$) and (iii) non-spherically symmetric explosion with complete lateral mixing ($Aspherical^\star$). In all cases is taken into account $M_{\rm He}=4 M_{\odot}$, $M_{2,i}=2M_{\odot}$ and $a_{c,i}=10R_{\odot}$. $Q$ and $Q'$ factors show the quality of the comparison between modeled and observed abundances for all and 6 elements (Mg, Si, S, Ti, Ni and Fe), respectively.\label{tabsn}}
\centering
\scalebox{1.05}[1.0]
{
\begin{tabular}[c c c c c c c c c c c c c c c ]{c c c c c c c c
c c c c c c c c c}
& & \multicolumn{12}{c}{[X/H]$_{\rm Expected}$}\\
\cmidrule{3-13} 
& &\multicolumn{11}{c}{$m_{\rm cut}(M_{\odot})$}\\
\cmidrule{3-13} 
& &\multicolumn{3}{c}{$Spherical$}
&\multicolumn{5}{c}{$Aspherical$}
&\multicolumn{3}{c}{$Aspherical^\star$}\\
	\cmidrule{3-5}\cmidrule{7-9}\cmidrule{11-13}
Element & [X/H]$_{\rm Obs}$ & 1.31 &1.34 & 1.36 &  & 1.28 & 1.36 & 1.45 & &1.49 & 1.55 & 1.77\\
		\cmidrule{1-13} 
		\cmidrule{1-13} 
	\multicolumn{8}{c}{            }\\

& &\multicolumn{3}{c}{$f_{\rm cap}=0.3$}
&\multicolumn{9}{c}{$f_{\rm cap}=0.1$}\\
\cmidrule{3-5}\cmidrule{7-13}
O & 0.07 & 0.17 & 0.17 & 0.17 && 0.45 & 0.45 & 0.42 & & 0.34 &
0.34 & 0.30 \\
Mg & 0.87 & 0.17 & 0.17 & 0.17 & & 0.47 & 0.47 & 0.44 & &0.34 &
0.34 & 0.32 \\
Al & 0.42 & 0.20 & 0.20 & 0.20 & & 0.52 & 0.52 & 0.50 && 0.38 &
0.38 & 0.36 \\
Si & 0.52 & 0.22 & 0.22 & 0.22 & & 0.56 & 0.39 & 0.30 & & 0.42 &
0.42 & 0.19 \\
S & 0.52 & 0.13 & 0.13 & 0.13 & & 0.40 & 0.14 & 0.08 & & 0.29 &
0.29 & 0.02 \\
Ca & 0.27 & 0.13 & 0.12 & 0.12 & &0.37 & 0.11 & 0.10 & & 0.33 &
0.29 & 0.00 \\
Sc & 0.67 & 0.02 & 0.00 & 0.00 && 0.02 & 0.00 & 0.00 & & 0.02 &
0.02 & 0.00 \\
Ti & 0.59 &0.21 & 0.18 & 0.16 && 0.17 & 0.09 & 0.05 & & 0.63 &
0.15 & 0.00\\
Fe & 0.27 & 0.21 & 0.20 & 0.17 & & 0.19 & 0.10 & 0.07 & & 0.33 &
0.17 & 0.00& \\
Ni & 0.42 &0.46 & 0.33 & 0.24 & & 0.15 & 0.10 & 0.04 && 0.44 &
0.14 & 0.00&\\
\cmidrule{3-13} 
& $Q$ &      2.71&      2.86& 3.03&     &      1.68&      2.55& 3.10&  &      1.44&      2.07& 4.23& \\
& $Q'$ &      4.89&      5.13& 5.43&      &      2.32&      3.82& 4.96&&      2.07&      3.42& 7.06&\\

	\multicolumn{7}{c}{            }\\

& &\multicolumn{3}{c}{$f_{\rm cap}=0.9$}
&\multicolumn{9}{c}{$f_{\rm cap}=0.2$}\\
\cmidrule{3-5}\cmidrule{7-13}
O & 0.07 & 0.38 & 0.38& 0.38 & & 0.67 & 0.66 & 0.63 & &0.53 &
0.53 & 0.48 & \\
Mg & 0.87 & 0.39 & 0.39& 0.39 && 0.69 & 0.69 & 0.66 && 0.53 &
0.53 & 0.50 & \\
Al & 0.42 & 0.44 & 0.44& 0.44 & & 0.75 & 0.75 & 0.72 && 0.58 &
0.58 & 0.56 &\\
Si & 0.52 & 0.47 & 0.47& 0.47 && 0.80 & 0.59 & 0.47 & & 0.63 &
0.63 & 0.33 & \\
S & 0.52 & 0.32 & 0.32& 0.32 & & 0.60 & 0.25 & 0.15 & & 0.46 &
0.46 & 0.04 & \\
Ca & 0.27 & 0.30 & 0.30& 0.30 & & 0.57 & 0.20 & 0.18 && 0.51 &
0.46 & 0.00& \\
Sc & 0.42 &0.07 & 0.01 & 0.01 && 0.04 & 0.00 & 0.00 & & 0.04 &
0.04 & 0.00 \\
Ti & 0.59 & 0.46 & 0.40& 0.37 & &0.30 & 0.16 & 0.09 & & 0.88 &
0.25 & 0.00 &\\
Fe & 0.27 & 0.46 & 0.43& 0.39 & & 0.33 & 0.19 & 0.13 & & 0.52 &
0.30 & 0.00 &\\
Ni & 0.42 & 0.82 & 0.65& 0.51 & & 0.26 & 0.18 & 0.07 && 0.65 &
0.24 & 0.00 &\\
		\cmidrule{1-13} 
& $Q$ &  1.79&      1.63& 1.48&&      1.84&      1.99& 2.52&&      2.02&      1.43& 3.70&\\
& $Q'$ &   2.39&      2.02& 1.88&&      1.40&      1.87& 3.04&&      1.72&      1.54& 5.49&\\
\end{tabular}
}
\end{table*}
\subsection{Aspherical SN explosion}
The spherically symmetric supernova explosion is not the only plausible scenario for this X-ray binary system. \citet{gon05a} studied both the spherical and non-spherical SN explosions to try to explain the observed abundances in the secondary star of the NSXB Centaurus X-4. They concluded that the spherical case produce better fits to the observed data than the aspherical model. This is consistent with the global picture in which a relatively low-mass primary star ($M_1 \sim  20 M_{\odot}$) may leave a neutron star and explode as a normal SN, with an energy of $\sim10^{51}$ ergs, as suggested in e.g. \citet{nom03}. A normal SN may be less aspherical than a hypernova, as proposed for the origin of the black hole in the BHXB Nova Sco 94 \citep[see e.g.][]{isr99,bro00,pod02}. 
However, \citet{gon08a} performed a detailed chemical analysis including more elements as Al and Ca, which are produced in outer and inner layers of the explosion respectively, and claimed that the best fit is obtained for an spherically symmetric more energetic hypernova model.  

Both Cen X-4 and Nova Sco 94 have large systemic velocities, $\gamma \sim 190$ km~s$^{-1}$~\citep{cas07}  and $ \gamma \sim -167$ km~s$^{-1}$~\citep{gon08a}, which may indicate a signature of an additional natal kick in an asymmetric supernova explosion. The systemic velocity of Cyg X-2 is very large, $\gamma \sim -210$ km~s$^{-1}$~\citep{cas10}. This may also point to an asymmetric supernova.  A symmetric mass ejection during the SN explosion of maximum mass values of the He core of $[4.8-6.3]$~$M_{\odot}$ and a secondary star of  $[2.0-3.5]$~$M_{\odot}$ provides a impulse~\citep[Blaauw-Boersma kick][]{bla61,boe61} for the system of $v_{\rm sys} \sim 145 - 215$ km~s$^{-1}$~\citep{nel99,bro01}. This could explain the large $\gamma$ velocity of the system only in the extreme case, although maybe possible, of $M_{\rm He}\sim 6.3$~$M_{\odot}$ and $M_{2,i}\sim 3.5$~$M_{\odot}$. However, for the case we are considering here, i.e. $M_{\rm He}\sim 4.0$~$M_{\odot}$ and $M_{2,i}\sim 2.0$~$M_{\odot}$, gives $v_{\rm sys} \sim 115$ km~s$^{-1}$, which may require additional natal kick from an asymmetric supernova.

We thus have also taken into account aspherical supernova explosions to study the case of Cygnus X-2. Aspherical explosions produce chemical inhomogeneities in the ejecta that could fit our observed values in a way spherically symmetric explosion models do not. An aspherical explosion produces nucleosynthetic inhomogeneities dependent on direction: thus, if the jet in the aspherical explosion is collimated perpendicular to the orbital plane, where the secondary star is located, elements such as Ti, Ni and Fe are ejected in the jet direction, while elements like O, Mg, Al, Si and S are ejected in the equatorial plane of the helium star \citep{mae02} and thus expected to be enhanced in the secondary star.

Following the procedure in the spherical case, we have considered two different asphericals models (depending of the direction of the jet) for a $\sim 16M_{\odot}$ progenitor with $M_{\rm He} = 4.0 M_{\odot}$ He core~\citep{mae02} and $M_{2,i}=2.0 M_{\odot}$. The expected abundances in the secondary star after being polluted by supernova products (assuming solar-abundance models) are also given in Table~\ref{tabsn}. For the first model, where most of the matter is ejected in the equatorial plane, best fits are shown in Fig.~\ref{figasn}, for $f_{\rm cap}=0.1$ (top panel) and $f_{\rm cap}=0.2$ (bottom panel). Three different mass cut values are also taken into account. 
In these cases, dependence on the mass cut is important for elements with Z greater than 14, this is for elements heavier than Al. The three different mass cut values in the range 1.5-1.8~$M_{\odot}$ seem to provide a good agreement with the observed abundances of the secondary star in Cyg X-2, and only requires that about 20\% (($f_{\rm cap} \leq 20\%$) of the ejected matter in the equatorial plane, i.e. the direction pointing to the secondary star, is captured. Contrary to the case of the spherical case, the Mg abundance can be approximately reproduced by the aspherical explosion, although Al and Sc are marginally reproduced in that case. 

In Fig.{\ref{figlsn} we display the expected abundances in the secondary star using the assumption of complete lateral mixing~\citep[see][for further details]{mae02,pod02}, where the ejected matter is completely mixed with each velocity bin \cite[see also][]{gon08a}. \citet{pod02} pointed out that the assumption of complete lateral mixing is extreme; at present we cannot identify any physical process that would lead to such a result. We can see that observed abundances are also well reproduced if we assume complete lateral mixing, but only at mass cut values below 1.55~$M_{\odot}$. These aspherical models also require low capture efficiencies.

The expected results for the solar abundance aspherical models (see Table~\ref{tabsn}), agree with our observed values, within errors bars, in most cases. However, some elements, such as Si, S, Ca, Fe and Ni, are strongly sensitive to the mass cut, and therefore, a slight change of this parameter can affect the expected values. The quality factors $Q$ and $Q'$ indicate that the best model that matches reasonably well the observed abundances is the aspherical model with a mass cut of about 1.3~$M_{\odot}$ and a capture efficiency of $\sim 20$~\%, although this model is unable to reproduce the observed oxygen abundance. In the case of complete lateral mixing, O and Mg abundances do not seem to have the same degree of agreement as other elements. 

\begin{figure}
\begin{center}
{\includegraphics[scale=0.5]{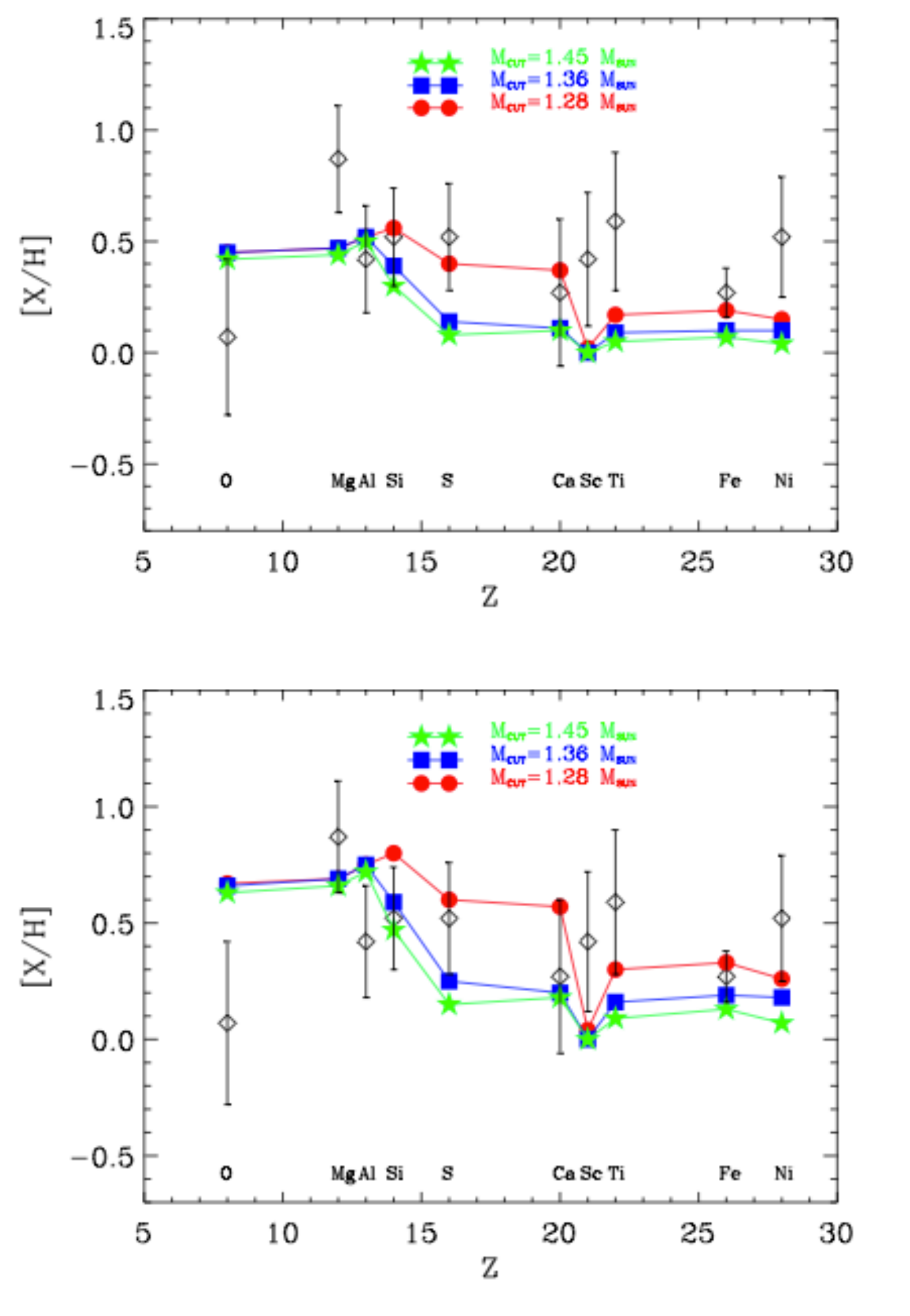}}
\caption{Expected abundances in the secondary atmosphere due to the pollution caused by a non-spherically symmetric core-collapsed SN explosion model, 
with a capture efficiency of $f_{\rm cap}=0.1$ (top panel) and $f_{\rm cap}=0.2$ (bottom panel). A kinetic energy $E_{K} = 1 \times 10^{51}$ergs and an orbital distance of $a_{c,i}=10R_{\odot}$ have been adopted. In this case, most of the matter is ejected in the equatorial plane of the primary, where the secondary star is expected to be located. Three mass cuts are depicted.\label{figasn}}
\end{center}
\end{figure}

\begin{figure}
\begin{center}
{\includegraphics[scale=0.5]{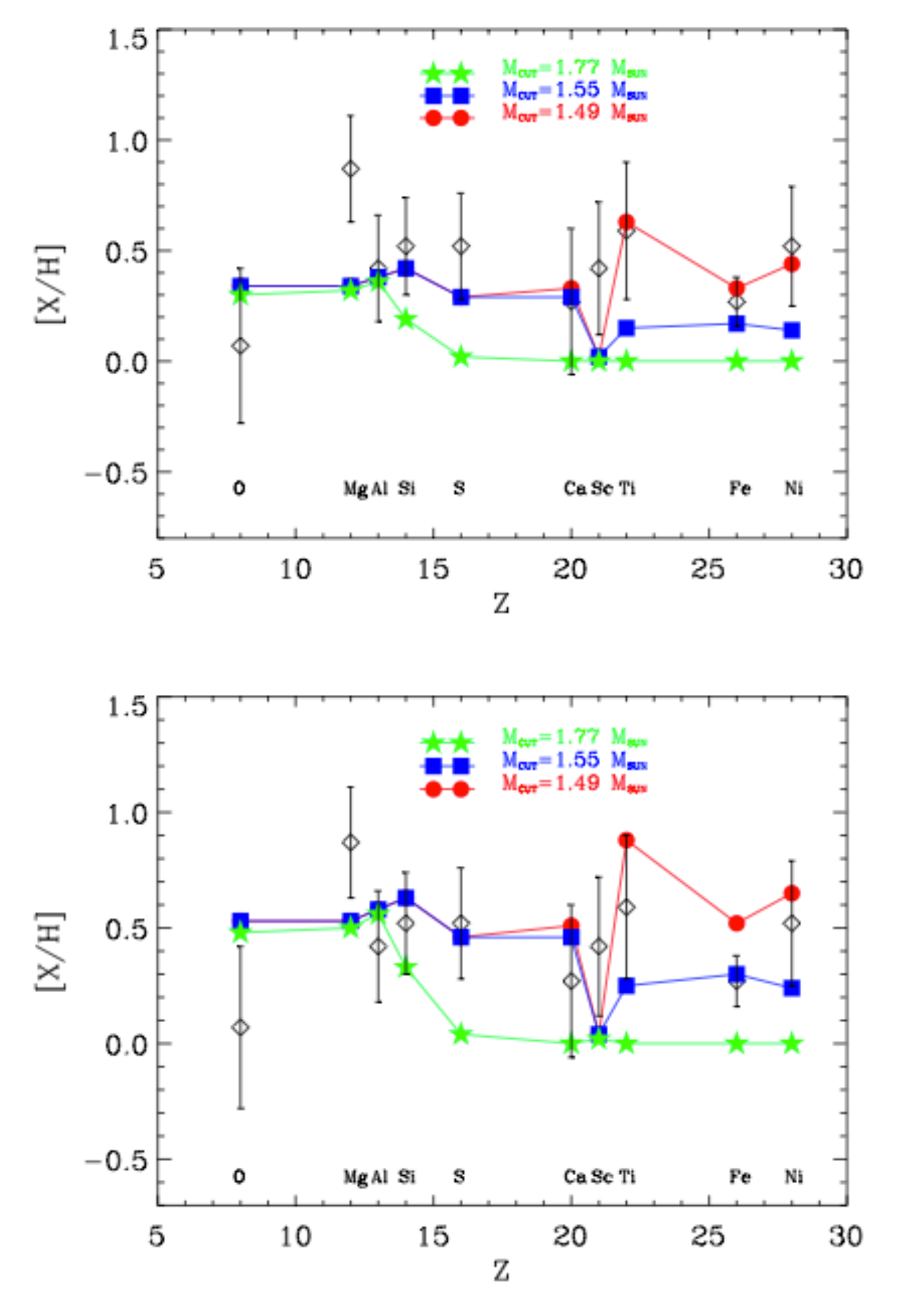}}
\caption{Same as Fig. 11, but assuming complete lateral mixing.\label{figlsn}}
\end{center}
\end{figure}

The unusual nature of the secondary star, as studied by \citet{king99,pod00}, already makes it difficult to assess the evolutionary status of the system.
This comparison using both spherically and aspherically symmetric SN models allow us to evaluate different scenarios but we cannot rule out any model. Taking into account the limitations in both the quality of the data and SN models available, and our simple model of contamination of the secondary star, we suggest an aspherical SN explosion of a $4 M_{\odot}$ He core as the origin of the neutron star in Cygnus X-2. 
This result seems to be the opposite as proposed in \citep{gon05a}, where a spherically symmetric SN explosion provided better agreement with the observed abundances. More massive progenitors of compact objects are expected to produce more energetic hypernovae~\citep{nom03,nom10} which may be more asymmetric than normal supernovae~\citep{mae06}. However, less massive progenitors may show, according to these authors, different levels of kinetic energies.
Our aspherical SN model has a given degree of asymmetry given by the SN model parameters provided in \citet{gon05a} that we cannot tune and seem to be the preferred model to explain the observed abundances. This may indicate that whereas hypernovae may tend to be more asymmetric, ``normal'' supernovae could explode at different symmetries.

The detection of high abundance of Fe in NSXBs may be a signature of the explosion that give rise to the formation of the NSs and the ejection of a significant amount of Fe, that was captured by secondary stars in these systems. However, this is somehow in contradiction with derived kinetic energies and the ejected $^{56}$Ni mass fractions from observations of recently detected supernovae~\citep[see e.g.][]{nom10}, where hypernova explosions that give rise to the formation of BHs from more massive progenitors are expected to eject larger amounts of $^{56}$Ni than less massive progenitors that produce NSs at lower kinetic energies and smaller amounts of  $^{56}$Ni.

\section{Conclusions}

We have presented the high-quality spectra of the secondary star in Cygnus X-2 and performed a detail chemical analysis of the atmosphere of the secondary star.
We have used a technique that allows determination of stellar parameters and metallicity, taking into account a possible veiling from the accretion disk. We find $T_{\rm eff}=6900 \pm 200$~K, $\log g = 2.80 \pm 0.20$, [Fe/H]$=0.27 \pm 0.19$, and a disk veiling (defined as $F_{\rm disk}/F_{\rm total}$) of about $55 \%$ at 5500~{\AA} and decreasing towards longer wavelengths.

The global metallicity of the secondary star is supersolar which may be a signature of the explosion that formed the neutron star in this system.
$\alpha$-elements like Mg, Si, S, Ti are strongly enhanced and Fe-peak elements like Ni are significantly over-abundant. Surprisingly the O abundance might be depleted in this star (uncertainties in the NLTE correction only allow us to consider an upper limit to the abundance). This low abundance may suggests diffusion effects, although recent models including atomic diffusion, radiative accelerations and turbulence are unable to explain the observed abundances, in particular, the relative enhancements of Mg, Si and S.

We have explored the possibility that the neutron star in Cygnus X-2 formed in a SN explosion and that the secondary star could have captured some of the ejected matter in the supernova.
We have compared the observed abundances with those expected in the possible scenario of pollution from SN nucleosynthetic products. We consider both spherically symmetric and non-spherically symmetric SN explosion models with several tunning parameters such as the mass cut and the fraction of matter captured from the ejecta.
Most of the abundance anomalies can be explained assuming a low mass cut in our SN models, at about $\sim1.4-1.6 M_{\odot}$, which allow elements such as Ni, Ti, S and Si to escape from the collapse and formation of the neutron star and therefore, to be able to pollute the secondary star.
 
The high systemic velocity of Cygnus X-2 ($\gamma \sim -210$~km~s$^{-1}$) may suggest an asymmetric mass ejection during the supernova explosion. Aspherical supernova models seem to better reproduce the observed abundances than spherically symmetric models. New high-resolution, high-signal-to-noise spectra of this NSXB at different wavelength ranges are encouraged to improve this abundance analysis and to measure more elements in the secondary star of Cygnus X-2.

\section*{Acknowledgements}

This work has made use of the VALD database and IRAF facilities
and has been partially financed by the Spanish Ministry projects
AYA2011-29060. 
J.I.G.H. also from the Spanish Ministry of Economy and 
Competitiveness (MINECO) under the 2011 Severo Ochoa Program 
MINECO SEV-2011-0187.
We thank S. Bertr\'an de Lis for providing us with her oxygen Galactic 
trends prior to publication.

\bibliographystyle{mn2e}

\label{lastpage}

\end{document}